\documentclass[showpacs,showkeys]{revtex4}

\usepackage{amssymb}
\usepackage{amsmath}
\usepackage{amsfonts}
\usepackage{graphicx}
\providecommand{\rbr}[1]{\left( #1 \right)} %
\providecommand{\sqbr}[1]{\left[ #1 \right]} %
\providecommand{\mt}[1]{\mathrm{#1}} %
\providecommand{\mc}[1]{\mathcal{#1}}%
\providecommand{\mb}[1]{\mathbb{#1}}%
\def\ra{\rightarrow}

\begin{document}
\title[ ]{The maximization of Tsallis entropy with complete deformed functions and the problem of constraints}
\author{Thomas Oikonomou}

\email{thoikonomou@chem.demokritos.gr}

\affiliation{Institute of Physical Chemistry, National Center for Scientific Research
``Demokritos", 15310 Athens, Greece}

\author{G. Baris Bagci}

\affiliation{Department of Physics, Faculty of Science, Ege
University, 35100 Izmir, Turkey}


\begin{abstract}

We first observe that the (co)domains of the $q$-deformed functions
are some subsets of the (co)domains of their ordinary counterparts,
thereby deeming the deformed functions to be incomplete. In order to
obtain a complete definition of $q$-generalized functions, we
calculate the dual mapping function, which is found equal to the
otherwise \textit{ad hoc} duality relation between the ordinary and
escort stationary distributions. Motivated by this fact, we show
that the maximization of the Tsallis entropy with the complete
$q$-logarithm and $q$-exponential implies the use of the ordinary
probability distributions instead of escort distributions. Moreover,
we demonstrate that even the escort stationary distributions can be
obtained through the use of the ordinary averaging procedure if the
argument of the $q$-exponential lies in (-$\infty$, $0$].

\end{abstract}
\eid{ }
\date{\today }
\startpage{1}
\endpage{1}
\maketitle

\section{Introduction}\label{intro}
%

Although Tsallis generalization of Boltzmann-Gibbs (BG)
thermostatistics \cite{Tsallis1988} has found many diverse fields of
application \cite{Tsnewbook}, there are some important issues in
need of clarification such as the definition of temperature
\cite{Abe1,Abe2,Abe3,Abe4,Hasegawa} and the associated central limit
theorem \cite{Tirnakli1,Tirnakli2,Grassberger,Pluchino}. One such
subject of debate is the role of constraints in the maximization
procedure of the Tsallis entropy \cite{Plastino1}. One choice of
constraints consists of adopting the ordinary probability
distributions, whereas a second choice (historically third choice)
is adopted by considering the so-called escort distributions
\cite{Tsnewbook}. Although many fundamental features of the
underlying thermostatistics such as the Legendre structure seem to
be preserved in both choices of constraints \cite{Plastino1}, it is
not easy to comprehend the need and the physical meaning of a new
definition of probability distribution (and constraint) i.e., escort
distribution.

There have been many attempts to choose between these two distinct
types of constraints. For example, one such attempt has been made by
considering the Shore-Johnson axioms in order to choose between
these two probability distributions \cite{Bagci}. However, it was later rejected
on the ground of irrelevance \cite{Abe5}. Some recent attempts along these lines
have been made concerning the stability of these distributions \cite{Hanel}. It
seems though that a final verdict is hard to reach through these
considerations, too \cite{Abe6,Lutsko}.

However, all these aforementioned attempts explicitly depend on the
$q$-deformed functions present in the definition of the Tsallis
entropy expression. In fact, the mathematical structure of the
$q$-thermostatistics is usually obtained by replacing the ordinary
logarithmic and exponential definitions by their $q$-deformed
counterparts. On the other hand, these deformed generalizations are
not \textit{complete} as their ordinary counterparts i.e., the
deformed (co)domains do not extend to the whole appropriate range
and are dependent on the deformation parameter $q$. This physically
implies that these deformed functions can map the arguments only to
a restricted codomain, not due to the constraints imposed by the
system, but due to the incompleteness of the deformed functions.
Moreover, due to the feature of incompleteness, these deformed
functions are not invertible in entire (co)domain. In other words,
the $q$-thermostatistics based on the incomplete deformed functions
is unable to explain the asymptotic inverse power law decays in a
consistent manner. Therefore, all discussions based on these
incomplete deformed functions are in difficulty of being incomplete,
too. This is also the case when one tries to discuss the nature of
constraints in the maximization procedure of the associated Tsallis
entropy.

In order to overcome the incompleteness of the $q$-deformed
functions, we propose the existence of a dual mapping $d(q)$ in such
a way that the deformed functions become invertible in the entire
domains of their ordinary counterparts. We also note that the
criterion of completeness based on finding a dual function $d(q)$ is
general (see Ref. \cite{Thomas} for details) and can be applied to
other deformed functions such as the ones by Kaniadakis
\cite{Kaniadakis} and Abe \cite{Abes}. Doing this, we also eliminate
the dependence of the intervals on the deformation parameter $q$.
This enables the only constraints to be the ones imposed by the
system. Then, we maximize the Tsallis entropy by using the complete
$q$-deformed functions. The main result of this complete
maximization procedure is to show that the escort distributions used
so far are merely a means to compensate for the incompleteness of
the underlying mathematical structure, and hence irrelevant to the
formalism, when one uses the complete deformed functions. The
ordinary probability distribution does not only suffice to maximize
the Tsallis entropy, but also proves to be the correct constraint to
be used in the maximization procedure. We also show how the escort
distribution emerges through ordinary averaging procedure, and
comment on its relevance in a consistent framework.

The paper is organized as follows. In Section II, we first
demonstrate the incompleteness of the $q$-deformed functions and
then define their complete counterparts. We maximize the Tsallis
entropy based on complete deformed functions to show that the escort
distributions are redundant in Section III.  Concluding remarks is
presented in Section IV.

\section{Complete $q$-generalized functions}\label{sec:2}

Tsallis generalization is based on the following definition
of deformed functions
%
\begin{equation}\label{Ts-functions}
f_{q}(x):=\frac{x^{1-q}-1}{1-q},\qquad\qquad
h_{q}(x):=\Big[1+(1-q)\,x\Big]^{\frac{1}{1-q}},
\end{equation}
%
with $h_q\equiv f^{-1}_q$, $f_{q\ra q_{0}\equiv1}(x)=\ln(x)$ and $h_{q\ra
q_{0}\equiv1}(x)=\exp(x)$.
The aforementioned functions are characterized by the following identities
%
\begin{equation}\label{Ts-identities}
f_{q}(x)+f_q(1/x)=\Theta_q(x)\qquad\quad\Leftrightarrow\quad\qquad
h_{q}(-x)h_q(x)=\Lambda_q(x),
\end{equation}
%
with $\Theta_q(x):=\frac{x^{1-q}+x^{q-1}-2}{1-q}$ ($x>0$) and $\Lambda_q(x):=\sqbr{1-(1-q)^2 x^2}_+^{1/(1-q)}$ where $[X]_+=\mt{max}\{X,0\}$.
The functions $\Theta_q$ and $\Lambda_q$ are related in the sense
that $\Theta_q(x)\ne0\;\Leftrightarrow\;\Lambda_q(x)\ne1$ or
equivalently $\Theta_q(x)=0\;\Leftrightarrow\;\Lambda_q(x)=1$.
Before proceeding further, we define the
following intervals
$\mathbb{L}_{0}:=\{x\in\mathbb{R}:0<x\leqslant1\}$,
$\mathbb{L}_{1}:=\{x\in\mathbb{R}:x\geqslant1\}$ so that
$\mathbb{L}_{0}\bigcup\mathbb{L}_{1}=$
$\mathbb{R}^+:=\{x\in\mathbb{R}:x>0\}$ i.e., the set of all positive
real numbers. We also have the intervals
$\mathbb{R}^+_0:=\{x\in\mathbb{R}:x\geqslant0\}$ and
$\mathbb{R}_0^-:=\{x\in\mathbb{R}:x\leqslant0\}$, respectively so
that $\mathbb{R}=\mathbb{R}^{-}_0\bigcup\mathbb{R}^{+}_0$ is the set
of all real numbers. Concerning the domain and codomain of the
$q$-functions in Eq. \eqref{Ts-functions}, we observe
%
\begin{subequations}\label{Ts-fun-(co)do}
\begin{align}
q<1& \qquad f_q:\mathbb{R}_0^+\longrightarrow
\bigg[\frac{1}{q-1},\infty\bigg),
\qquad\quad\,%
h_q:\bigg[\frac{1}{q-1},\infty\bigg)\longrightarrow\mathbb{R}^+_0,\\
q=1& \qquad f_q:\mathbb{R}^+\longrightarrow \mathbb{R},
\qquad\quad\quad\;\,\,\quad\qquad%
h_q:\mathbb{R}\longrightarrow\mathbb{R}^+,\\
q>1& \qquad f_q:\mathbb{R}^+\longrightarrow
\bigg(-\infty,\frac{1}{q-1}\bigg],
\qquad%
h_q:\bigg(-\infty,\frac{1}{q-1}\bigg]\longrightarrow\mathbb{R}^+.
\end{align}
\end{subequations}
%
Eq. \eqref{Ts-fun-(co)do} shows that the functions
$f_{q}$ and $h_{q}$ represent a possible generalization only in
some subsets of the codomains of the ordinary logarithmic and
exponential functions when $q\ne1$.
In other words, these $q$-deformed functions are not invertible in the entire (co)domains
of their ordinary counterparts. Since they are not generally
invertible, a genuine $q$-exponential decay is ill-defined.
Furthermore, from Eq. \eqref{Ts-identities}, we see that the
function $h_q(-x)$ does not describe a $q$-exponential decay
$1/h_q(x)$, as has been frequently considered in literature.
Therefore, we call these functions as incomplete generalized
functions.
Searching for the origin of the aforementioned incompleteness we inspect e.g., the
ordinary logarithm (in natural base $e$). We observe that
$\ln:\mathbb{R}^+\ra\mathbb{R}$ has an argument intercept at the
point $x=1$, dividing the logarithmic domain $\mathbb{R}^+$ into two
subdomains $\mathbb{L}_0$ and $\mathbb{L}_1$. The relation between
the images of the log-arguments $x\in\mathbb{L}_1$ and
$1/x\in\mathbb{L}_0$ is represented by
%
\begin{align}
\label{ord-log}%
f_{q_0}(x)+f_{q_0}(1/x)=0.
\end{align}
%
Since the right hand side of Eq. \eqref{ord-log} is equal to zero,
the logarithmic codomain for $x\in\mb{R}^+$ is $\mb{R}$. If the sum
of the above logarithmic terms would be equal to $\Theta(x)$,
$\Theta$ being a real non-singular function, then the codomain of
the logarithmic function would depend on the codomain of $\Theta(x)$
as well i.e., $\ln:\mb{R}^+\ra\mb{M}\subseteq\mb{R}$, where $\mb{M}$
is some subset of the set of all real numbers. Moreover, inverting
Eq. \eqref{ord-log}, one would obtain $\exp(-x)\ne1/\exp(x)$ for
$\Theta(x)\ne0$.
It becomes then evident that the origin of the incompleteness lies on the existence of the functions $\Theta_q$ and $\Lambda_q$.
Due to this incompleteness, $q$-thermostatistics is bound to accept
arguments and yield results only in limited (co)domains.
Furthermore, in these (co)domains the deformed exponential function
$h_q$ exhibits different features for positive or negative
arguments.

In order to define complete $q$-generalized functions, one must try
to equate $\Theta_{q}(x)$ to zero (or equivalent $\Lambda_q$ to
unity). Before proceeding further, let us denote the range of
validity for the parameter $q$ by $\mc{A}_{q}\subseteq\mathbb{R}$. A
boundary value of $\mc{A}_{q}$ is always $q_0=1$ e.g.,
$\mc{A}_{q}:=(\alpha,1]$ with $\alpha<1$ or $\mc{A}_{q}:=[1,\alpha)$
with $\alpha>1$. For defining complete deformed functions, we assume
the existence of a function $d(q):\mc{A}_{q}\ra\mc{B}_{q}$
($\mc{B}_{q}\subseteq\mathbb{R}$) such that

\begin{align}\label{NewRel1a}
f_{q}(x)+f_{d(q)}(1/x)&=f_{q}(1)
\end{align}
%
with $\lim_{q\rightarrow 1}d(q)=1$. Setting $x=1$ in Eq.
\eqref{NewRel1a}, we obtain $f_{d(q)}(1)=0$ in accordance with the
ordinary logarithm. It can be seen that the above relation is valid
for any set of parameters by substituting $q'=d(q)$ into the above
equation. Therefore, Eq. \eqref{NewRel1a} can be rewritten as

\begin{align}\label{x-corresp}
f_{q}(x)&=-f_{d(q)}(1/x)
\end{align}
%
in analogy to Eq. \eqref{ord-log}. The mapping $d(q)$ is called as
the dual function and the correspondence $q\leftrightarrow d(q)$ as
duality relation.

We further observe that the argument $x=1$, since $f_{q}(1)=0$,
divides the (co)domain $\mathbb{R}^+$ ($\mathbb{R}$) into the
following sub(co)domains
$\mathbb{R}^+=\mathbb{L}_0\bigcup\mathbb{L}_1$
($\mathbb{R}=\mathbb{R}^-\bigcup\mathbb{R}_0^+$). Consequently, we
are led to the complete $q$-deformed definitions
%
\begin{align}\label{(co)domain2-a}%
\ln_{q}:=\begin{cases}
f_{d(q)}:\mathbb{L}_0\longrightarrow\mathbb{R}^{-}_0\\
\\
f_{q}:\mathbb{L}_1\longrightarrow\mathbb{R}^{+}_0
\end{cases}
\qquad\mt{and}\qquad
\exp_{q}\equiv\ln_{q}^{-1}:=
\begin{cases}
h_{d(q)}:\mathbb{R}^{-}_0\longrightarrow\mathbb{L}_0\\
\\
h_{q}:\mathbb{R}^{+}_0\longrightarrow\mathbb{L}_1
\end{cases}\,,
\end{align}
%
with $q\in\mc{A}_{q}$.
We can then, for the complete generalized logarithm and exponential
given above, verify that
%
\begin{align*}
\ln_{q}(1/x)&=f_{d(q)}(1/x)=-f_{q}(x)=-\ln_{q}(x),\\
\exp_{q}(-x)&=h_{d(q)}(-x)=\frac{1}{h_{q}(x)}=\frac{1}{\exp_{q}(x)}
\end{align*}
%
for $x\in\mb{L}_1$ and $x\in\mb{R}^+_0$, respectively.
%
%

The parameter range $\mc{A}_{q}$ can be determined by requiring the
fulfillment of the following limits satisfied by the ordinary
functions
%
\begin{subequations}\label{sec2-eq:C}
\begin{align}
\label{GenCon1}%
\lim_{x\ra0}\exp_{q}(x)&=1, &\lim_{x\ra1}\ln_{q}(x)&=0,\\
\label{GenCon2}%
\lim_{x\ra-\infty}\exp_{q}(x)&=0,&\lim_{x\ra0}\ln_{q}(x)&=-\infty,\\
\label{GenCon3}%
\lim_{x\ra\infty}\exp_{q}(x)&=\infty,&\lim_{x\ra\infty}\ln_{q}(x)&=\infty,\\
\label{GenCon4}%
\lim_{x\ra-\infty}\frac{d}{dx}\exp_{q}(x)&=0,&\lim_{x\ra\infty}\frac{d}{dx}\ln_{q}(x)&=0.
\end{align}
\end{subequations}

Condition \eqref{GenCon1} ensures the continuity of $\ln_{q}(x)$ and
$\exp_{q}(x)$ at the points $x=1$ and $x=0$, respectively. The
behavior of the complete deformed functions at the boundaries of
their domains is determined from Eqs. \eqref{GenCon2} and
\eqref{GenCon3}. The condition \eqref{GenCon4} allows one to
preserve the same absolute maximum and minimum values of the
ordinary functions.

Having obtained the criterion of completeness in detail, we see that
the first step is to calculate the dual function $d(q)$. We can
calculate it from Eqs. \eqref{Ts-functions} and \eqref{x-corresp} as

\begin{align}\label{Ts-DualRel}
d(q)&=2-q.
\end{align}
%

The requirements listed in Eq. \eqref{sec2-eq:C} confines the values
of the deformation parameter $q$ into the following interval

\begin{equation}\label{q-interval-a}
\mc{A}_q:=(0,1].
\end{equation}
%
The image of $\mc{A}_q$ under the dual mapping $d(q)$ i.e.,
$\mc{B}_q$ can be calculated from Eqs. \eqref{Ts-DualRel} and
\eqref{q-interval-a} as

\begin{equation}\label{q-interval-b}
\mc{B}_q:=[1,2).
\end{equation}

Having explicitly obtained the dual mapping function $d(q)$ and the
range of validity of the deformation parameter $q\in\mc{A}_q$, we
can now write the analytical expression of the complete
$q$-generalized functions
%
\begin{align}\label{Ts-fun-2}%
\ln_{q}(x):=\begin{cases}
\displaystyle\frac{x^{q-1}-1}{q-1}, & x\in\mb{L}_0\\
\\
\displaystyle\frac{x^{1-q}-1}{1-q}, & x\in\mb{L}_1
\end{cases}\,,
\qquad\qquad
\exp_{q}(x):=
\begin{cases}
\Big[1+(q-1)\,x\Big]^{\frac{1}{q-1}}, & x\in\mb{R}_0^-\\
\\
\Big[1+(1-q)\,x\Big]^{\frac{1}{1-q}}, & x\in\mb{R}_0^+
\end{cases}\,,
\end{align}
%
%
in accordance with Eq. \eqref{(co)domain2-a}. We note that a
slightly different expression for $\exp_{q}$ is also obtained in
Ref. \cite{Teweldeberhan} in the context of cut-off prescriptions
associated with the $q$-generalized exponential. The division of the
(co)domains yielding to a complete definition of the deformed
functions was first noticed in Ref. \cite{Teweldeberhan}. However,
this observation has not been pursued further as a criterion of
completeness therein. Furthermore, the current definition is in
accordance with the trace-form entropic definition based on $f_q$
(see next Section).

The summary of the above results can be provided in a rather compact
form below, considering two parameter ranges, $\mc{A}_q$ and
$\mc{B}_q$, and keeping a single expression in the entire
(co)domains instead i.e.,

\begin{align}\label{Ts-fun-3}
\ln_q\equiv f_q&:    \substack{x\in\mb{L}_0 \\
q\in[1,2)}\bigcup\substack{x\in\mathbb{L}_1\\ q\in(0,1]}
\longrightarrow \substack{x\in\mathbb{R}^-_0\\
q\in[1,2)}\bigcup\substack{x\in\mathbb{R}_0^+\\ q\in(0,1]},
\qquad%
\exp_q\equiv h_q:\substack{x\in\mathbb{R}_0^-\\
q\in[1,2)}\bigcup\substack{x\in\mathbb{R}_0^+\\ q\in(0,1]}
\longrightarrow \substack{x\in\mathbb{L}_0\\
q\in[1,2)}\bigcup\substack{x\in\mathbb{L}_1\\ q\in(0,1]}.
\end{align}

\section{The maximization of the Tsallis entropy: ordinary versus escort distributions}\label{sec:3}

The Tsallis entropy expression reads

\begin{equation}\label{tsallis-entropy}
S_{q}(p_i):=\frac{\sum_{i=1}^{\Omega q}p_{i}^{q}-1}{1-q},
\end{equation}

where $p_i$ is the probability of the $i$th-configuration and
$\Omega_q$ is the maximum configuration function of the respective
$q$-ensemble. The above expression can be rewritten in terms of the
deformed $q$-logarithm given by Eq.  \eqref{Ts-functions} as

\begin{equation}\label{tsallis-entropy2}
S_{q}(p_i)=\sum_{i=1}^{\Omega_q}p_i f_{q}\rbr{1/p_i}.
\end{equation}

In the literature of $q$-thermostatistics, there is no consensus on
how the Tsallis entropy must be maximized. There are two distinct
choices for the constraints used throughout the literature. The
first one is based on considering the ordinary probability
distributions when one averages the constraints in the functional to
be maximized. This can be written as

\begin{equation}\label{maximization-ordinary}
\delta \left( S_{q}(p_i)-\alpha \sum\limits_{i}p_{i}-\beta
\sum\limits_{i}p_{i}\varepsilon _{i}\right) =0.
\end{equation}

The maximization of this functional yields the stationary
distributions, denoted by $\widetilde{p}_{i}$, of the form

\begin{equation}\label{ordinary-solution}
\widetilde{p}_{i}=\frac{1}{h_{q}(\varepsilon _{i})}.
\end{equation}

Second choice rests on the definition of escort distributions. The
escort distributions $P_{i}$ are defined as

\begin{equation}\label{escort}
P_{i}=\frac{p_{i}^{q}}{\sum\limits_{k}p_{k}^{q}}.
\end{equation}

The maximization of the Tsallis entropy with the escort
distributions becomes

\begin{equation}\label{maximization-escort}
\delta \left( S_{q}(p_i)-\alpha \sum\limits_{i}P_{i}-\beta
\sum\limits_{i}P_{i}\varepsilon _{i}\right) =0.
\end{equation}

The above maximization yields the following stationary distribution

\begin{equation}\label{escort-solution}
\widetilde{p}_{i}=\frac{1}{h_{2-q}(\varepsilon
_{i})}=h_{q}(-\varepsilon _{i}).
\end{equation}

However, both of these results rely on the use of Tsallis entropy
$S_{q}$ and this entropy in turn is based on the incomplete
$q$-logarithm. Considering that one uses the complete deformed
functions described above, we can have a more consistent look into
the nature of stationary distributions obtained in the maximization
procedure.

The first observation considers the argument of the $q$-logarithm
used in the definition of the Tsallis entropy. The microstate
probabilities vary between 0 and 1 i.e., $p_i\in[0,1]$ so that
$1/p_i$ will be equal to or greater than 1. This means that the
argument of the complete $q$-logarithm takes values in
$\mathbb{L}_1$. Then, the associated generalized statistics is
characterized by the deformation parameter range $\mc{A}_q=(0,1]$
according to Eq. \eqref{Ts-fun-3}. This parameter range corresponds
to the $q$-exponential from $\mathbb{R}_0^+$ to $\mathbb{L}_1$,
implying that the respective generalized exponential decay is
described by the function $1/h_q(x)$. Inspecting Eq.
\eqref{ordinary-solution}, we see that this corresponds to the
stationary distribution obtained from the maximization of the
Tsallis entropy with ordinary constraints.

Since the above considerations show that the maximization of Tsallis
entropy must be carried out with ordinary probability definitions,
one might wonder how the escort distributions emerge at all. In
order to shed light on this issue, we inspect Eqs. \eqref{Ts-fun-2},
\eqref{Ts-fun-3} and \eqref{escort-solution} to see that the
stationary probability distribution obtained from the constraints
averaged with the escort distribution corresponds to the parameter
validity range $q\in[1,2)$ and thus it should be related to the
expressions transformed under the mapping $d(q)\in\mc{B}_q$. This
parameter range corresponds to the complete $q$-exponential when
$x\in\mathbb{R}_0^-$ i.e., when the argument of the $q$-exponential
equal to or less than 0. Moreover, if we write the Tsallis entropy
compatible with this range i.e.,
\begin{equation}\label{d-transformedsolution}
S_{d(q)}(p_i)=\sum\limits_{i=1}^{\Omega _{d(q)}}p_{i}f_{d(q)}(1/p_{i})
           =-\sum\limits_{i=1}^{\Omega _{d(q)}}p_{i}f_{q}(p_{i})
           =\sum\limits_{i=1}^{\Omega_{2-q}}\frac{p_{i}^{2-q}-1}{q-1}
\end{equation}
and normalize it with ordinary constraints, we see that it
yields the stationary distribution given by Eq.
\eqref{escort-solution} i.e.,
$\widetilde{p}_{i}=1/h_{2-q}(\varepsilon _{i})$. This is the
stationary solution one obtains using the constraints averaged with
escort distributions.

To summarize, both types of stationary  distributions, ordinary and
escort, can obtained through the use of ordinary constraints.
However, it is the stationary distribution obtained through the
ordinary averages, which is consistent with the range of the
arguments, since the argument of the $q$-logarithm i.e., $1/p_i$, is
equal to or greater than 1. In other words, the escort distributions
are only \textit{ad hoc} means of conforming the incomplete
$q$-thermostatistics to the parameter range lying between 1 and 2.
Moreover, it is worth mentioning that this result is in agreement
with the concept of the entropy extensivity. Indeed, considering
Tsallis entropy in Eq. \eqref{tsallis-entropy2} for equal
probabilities $\widetilde{p}_i=1/\Omega_q$ and demanding the
fulfillment of the extensivity property with respect to the variable
$\varepsilon_i$ we are able to determine the maximum configuration
function given by
$\Omega_q(\varepsilon_i)=h_q(\varepsilon_i)\;\Rightarrow\;\widetilde{p}_i=1/h_q(\varepsilon_i)$.

In contrast to the incomplete definitions, both ordinary and escort
stationary distributions becomes identical in the context of the
complete definitions presented herein, since
$\exp_q(-x)=1/\exp_q(x)$, and can be obtained through ordinary
averaging procedure.

\section{Conclusions}\label{Concl}

The $q$-thermostatistics is fundamentally based on $q$-deformed
functions i.e., $q$-logarithm $f_q$ and $q$-exponential $h_q$.
Therefore, it is essential to use valid definitions of these
generalized functions. However, the aforementioned functions are not
bijective in the entire (co)domain of their ordinary counterparts,
implying that they do not represent complete generalizations, since
they are not be invertible in the entire (co)domain. This
incompleteness creates mathematical (and physical) discrepancies,
which makes the use of primary $q$-functions deficient. On the other
hand, the maximization procedure of Tsallis entropy necessitates the
use of the $q$-logarithm in its definition and its inverse i.e.,
$q$-exponential in obtaining the stationary solution resulting from
this maximization procedure. Therefore, it is important to
remaximize the Tsallis entropy with the corresponding complete
definitions of the generalized functions.

The maximization of Tsallis entropy in terms of the complete
$q$-generalized functions shows that the correct stationary solution
is the one associated with the ordinary constraint instead of the
widely used escort stationary distributions. In this sense, the
definition of the escort distributions is redundant.

It is worth noting that the use of $q$-generalized functions $f_q$
and $h_q$ is not solely limited to the entropy maximization
procedure. The inspection of these functions in a more ``complete"
framework shows that the escort distributions are nothing but
\textit{ad hoc} means of extending the so far incomplete definitions
of the deformed functions to the whole range of the deformation
parameter. In other words, the escort distributions become present
whenever the incomplete primary deformed functions cannot map
certain regions in the deformation parameter space $q$, namely,
$q\in[1,2)$. This region is physically important. Nevertheless, once
we define the complete $q$-deformed functions, all ranges of the
deformation parameter can be accessed in an invertible manner by
these complete functions, thereby showing the redundancy of the
escort distributions in $q$-thermostatistics.

Last but not least, all the definitions of the $q$-thermostatistics
based on the deformed functions must be revised accordingly in terms
of complete deformed functions. One such important example is the
recently defined $q$-Fourier transform \cite{Estrada}, since it is
closely related to obtaining a (possible) central limit theorem,
whose basin of attraction is $q$-Gaussian. We finally note that the
recently developed maximization procedure based on the definition of
the $q$-Fourier transform must be revised accordingly too in order
to assess the true role of the escort distributions \cite{Bagci2}.

\section*{Acknowledgments}
\noindent We thank U. Tirnakli for a careful reading of the
manuscript. GBB was supported by TUBITAK (Turkish Agency) under the
Research Project number 108T013.


\end{document}